\documentclass[lettersize,journal]{IEEEtran}

\usepackage[T1]{fontenc}
\usepackage{amsmath}
\usepackage{amssymb}
\usepackage{graphicx}
\usepackage{booktabs}
\usepackage{multirow}
\usepackage{array}
\usepackage{algorithm}
\usepackage{algorithmic}
\usepackage{url}
\usepackage{cite}
\usepackage{orcidlink}
\hypersetup{hidelinks}
\newtheorem{proposition}{Proposition}

\begin{document}
\bstctlcite{IEEEreferencecontrol}

\title{DScale: Scaling Block-Diffusion\\Speculative Decoding with Adaptive Verification}

\author{Rongjian~Chen\textsuperscript{\orcidlink{0009-0009-1091-2752}},~
Minxian~Xu\textsuperscript{\orcidlink{0000-0002-0046-5153}},~
\IEEEmembership{Senior~Member,~IEEE},\\
Zhengxin~Fang\textsuperscript{\orcidlink{0000-0003-2747-5199}},~\IEEEmembership{Graduate~Student~Member,~IEEE},\\
Kejiang~Ye\textsuperscript{\orcidlink{0000-0001-6133-407X}},~
\IEEEmembership{Senior~Member,~IEEE},~
and~Chengzhong~Xu\textsuperscript{\orcidlink{0000-0001-9480-0356}},~\IEEEmembership{Fellow,~IEEE}%
\thanks{R. Chen, M. Xu, and K. Ye are with the Shenzhen Institutes of
Advanced Technology, Chinese Academy of Sciences, Shenzhen 518052, China,
and also with the University of Chinese Academy of Sciences, Beijing 100049,
China (e-mail \{rj.chen2, mx.xu, kj.ye\}@siat.ac.cn).}
\thanks{Z. Fang is with Victoria University of Wellington, Wellington, New Zealand.}
\thanks{C. Xu is with the Institute of AI and Brain Sciences, Department of
Computer Science, University of Macau, Macau 999078, China (e-mail
czxu@um.edu.mo).}
\thanks{Corresponding author: Minxian Xu (e-mail mx.xu@siat.ac.cn).}
\thanks{Supplementary material accompanies this submission.}}

\markboth{IEEE Transactions on Computers}%
{Chen \MakeLowercase{\textit{et al.}}. Scaling Block-Diffusion Speculative Decoding
with Adaptive Verification}

\maketitle

\begin{abstract}
Growing large language model applications demand efficient inference.
At high concurrency, block-diffusion speculative decoding suffers from verification padding, rejected
candidates, and incompatibility between variable prefixes and fixed-shape graphs.
Uniform truncation sacrifices acceptable tokens.
We present DScale, preserving drafter architecture, weights, and full draft length.
A separate 112K-parameter predictor requires neither confidence calibration nor
hardware speed-curve preparation. Path-aware tiles reduce padding.
Dynamic verify-length (DVL) allocation packs scored prefixes into half the native
verification capacity. Fixed-address workspaces propagate changing boundaries
through verification and acceptance while reusing captured graphs.
On A100-40GB with tensor parallelism 1, Qwen3-8B and Qwen3-4B cover four datasets
and concurrency 8--32, reusing each target's frozen predictor.
Geometric-mean throughput gains across these configurations are respectively
43.9\% and 48.8\% over DFlash, 22.2\% and 37.7\% over DSpark,
and 24.4\% and 32.0\% over Domino, with lower request latency.
Cumulative ablations show that adding the three mechanisms successively
increases geometric-mean throughput, while budget adjustment improves
accepted-token retention. GPU profiling shows that complete decode-step time
on GSM8K decreases by 30.8--52.5\% relative to DFlash.
\end{abstract}


\section{Introduction}\label{sec:intro}
\IEEEPARstart{S}{erving} large language models (LLMs) under concurrent demand
requires low latency and high throughput~\cite{kwon2023efficient,sglang}.
However, LLMs commonly use autoregressive decoding, executing one full
model forward per token. This sequential execution limits generation speed.

\emph{Speculative decoding} uses a small \emph{draft} model to generate
candidates that the larger \emph{target} model verifies in
parallel~\cite{leviathan2023fast,chen2023accelerating}.
Accepting the longest prefix consistent with the target distribution
produces multiple tokens per forward when drafting is
accurate.
EAGLE and Medusa improve candidate quality through target-feature reuse and
tree-structured drafting~\cite{li2024eagle,cai2024medusa,li2024eagle2,li2025eagle3}.
Using block diffusion~\cite{arriola2025block}, DFlash conditions a lightweight
drafter on fused hidden features from multiple target layers, injected into
each draft layer's KV cache. Given this context and the last verified token,
it generates a complete block in one parallel forward pass for target
verification~\cite{chen2026dflash}.
At low concurrency, spare GPU capacity can absorb extra verification work.
However, target verification dominates GPU-kernel time in our Qwen3-8B
profile at concurrency~16, detailed in Section~\ref{sec:bg}.
DSpark combines a purpose-trained semi-autoregressive drafter with
confidence calibration and hardware-profiled scheduling~\cite{dspark}.
Target verification replays the smallest pre-captured token-capacity bucket
covering the batch's selected tokens.
Reducing verification work must still preserve accepted tokens.
Here, accepted tokens are newly drafted tokens retained by the target's
acceptance rule as a contiguous prefix, excluding the known anchor.

\begin{figure}[!t]
\centering
% Three-row source: DFlash_three_challenges_regular_rows_c3_arrows_20260911.pptx,
% slide 1; synchronized to intro_challenges.pptx/PDF on 2026-09-14.
\includegraphics[width=\columnwidth]{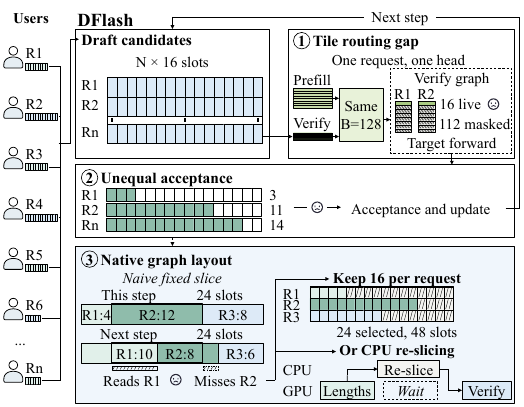}
\caption{Three connected challenges in DFlash serving.
(1) A shared prefill and verify tile leaves 16 live and 112 masked rows per request and head.
(2) Equal 16-slot windows accommodate unequal accepted prefixes.
(3) Changed boundaries make old slices read R1 and miss R2.
Equal-width padding restores 16 slots per request, while CPU re-slicing delays verification.
The dashed arrow links Challenges 2 and 3. Token lengths are illustrative.}
\label{fig:dflashintro}
\end{figure}

Fig.~\ref{fig:dflashintro} follows DFlash from request arrival through drafting,
verification, and acceptance. We retain the selected DFlash drafters'
training-time block configuration, reserving $W{=}16$ target-verification slots
per request for one known anchor and 15 new candidates.
Equal widths but unequal accepted prefixes expose three connected challenges.

\textbf{Challenge 1: Workload-shaped tiles.} A tile groups query rows for
attention. Prefill processes long input sequences, whereas verification
processes short candidate blocks. Sharing their tile layout leaves most
verify positions as inactive padding.

\textbf{Challenge 2: Request-varying acceptance.} Requests differ in how many
drafted tokens they will accept, but the accepted prefix is unknown until the
target verifies the candidates. A fixed width cannot match these request-level
differences.

\textbf{Challenge 3: Dynamic prefixes under a fixed verification-graph interface.}
Request-specific lengths move packed boundaries between steps.
Padding them back to DFlash's equal-width interface restores wasted work.
CPU re-slicing adds a host dependency.

Early-stopping methods adapt autoregressive draft length~\cite{specdecpp,adaedl},
whereas DFlash generates a full block in one parallel forward without
token-by-token stopping~\cite{chen2026dflash}. We study serving the same batch
with reduced candidate-slot capacity per verification graph while preserving
the drafter's architecture, weights, and full draft length. \emph{The problem is to
allocate this smaller capacity among request-specific candidate prefixes
according to acceptance potential, preserving generation progress while
carrying changing request boundaries through compact verification execution
and captured-graph replay.}

To address these challenges, we present \textbf{DScale}, a \emph{draft-training-free} runtime for
high-concurrency block-diffusion speculative decoding.
It reuses DFlash's drafter, retaining its architecture, weights, and full draft
length, without offline confidence calibration or hardware speed-curve profiling.
A 112K-parameter predictor trained once per target--drafter pair allocates
prefixes within half of each verification graph's native candidate-slot capacity.

Our contributions are as follows.
\begin{itemize}
\item We design path-aware tile routing to select precompiled attention
tiles for prefill and verification, removing padding beyond the verify
width while preserving prefill reuse.
\item We develop dynamic verify-length (DVL) budget allocation.
Predictor scores guide per-request prefix selection under a halved shared
budget, and flat-pack compaction preserves long useful prefixes without
shortening draft generation.
\item We integrate decision and verification graphs through a fixed-shape
compact workspace with dynamic request boundaries.
Both graphs share prefixes and boundaries through acceptance, preserving
compact execution without fixed-width padding or CPU re-slicing.
\item Across four datasets, two target models, and concurrency 8--32,
DScale improves geometric-mean throughput by 43.9--48.8\% and reduces
aggregate mean E2E latency by 31.4--33.1\% over DFlash.
Cumulative ablations show geometric-mean throughput rising with each added mechanism.
Budget adjustment raises accepted-token retention from about 71\% to over 91\%.
On GSM8K, GPU profiling shows 30.8--52.5\% shorter complete decode steps
than DFlash. Prediction, budget allocation, and packing together account for
3.08--4.08\% of step time in GPU execution at $c=8$.
\end{itemize}

\section{Background and Motivation}\label{sec:bg}

DFlash starts with target prefill to establish the prompt context.
At each decode step, a block-diffusion drafter uses the context and a known
anchor to propose 15 candidates in one parallel forward.
The anchor and candidates occupy \(W{=}16\) verification slots.
One target forward evaluates the block. Acceptance retains the longest
valid candidate prefix and produces a successor token at the first rejection
or after all candidates pass. The successor becomes the next anchor,
and drafting resumes with the updated context until the request
finishes~\cite{chen2026dflash}.
The native verification profile uses Qwen3-8B with a five-layer, approximately
1B-parameter DFlash drafter.

% Source: teacher_review_transfer_20260910/second_revision/native_graph_attribution.json.
Nsight Systems~\cite{nvidia2024nsight} profiling of native DFlash on
GSM8K, A100, TP1, and concurrency~16 attributes 83.3\% of summed GPU-kernel
time to the target verify graph, 10.9\% to the draft graph, and 5.7\% to
kernels outside these graphs. We collect 700 decode steps after 40 warmup
steps, sum kernel durations by graph launch excluding host time, and show that
verification is dominant and exposes three distinct bottlenecks requiring
different remedies.

\subsection{Verify Inherits Prefill's Tile}\label{sec:motiv:c1}

As illustrated by Challenge~1 in Fig.~\ref{fig:dflashintro},
native DFlash uses the same attention tile configuration for prefill and
speculative verification despite their different query lengths.
Prefill can supply thousands of query
rows per request, allowing a wide tile to reuse loaded KV data across rows.
Verification supplies only a short candidate block.
The native selector chooses a compile-time tile from hardware and head
dimensions, without distinguishing these two workloads.

Batch growth leaves this mismatch intact. Each computation block handles
one request's query-row tile for one head.
Adding requests increases the number of blocks without filling their unused query rows.
Distinct KV contexts and causal boundaries prevent requests from simply sharing a tile.
In the evaluated A100 configuration, the native tile has 128 rows, whereas
each verification request supplies $W{=}16$.
Thus 112 rows, or 87.5\%, are masked.
Masking excludes invalid rows from attention but does not compact the
compiled tile or resize its intermediate arrays.

DSpark's confidence scheduler selects candidates for verification~\cite{dspark},
but does not resize the compiled attention tile.
In the shared kernel above, shorter prefixes reduce active query rows
but leave vacant rows within the tile.

\textbf{\emph{Opportunity. Explore path-aware attention granularity based on
the different query lengths of prefill and speculative verification.
Match short verification tiles to the candidate width to reduce inactive
query-row positions, while retaining wide-tile KV reuse for prefill.
The two paths can then use layouts suited to their respective workloads
rather than share one layout.}}

\subsection{Request-Level Acceptance Variability and Verification Budget Allocation}\label{sec:motiv:c2}

Verification processes even rejected candidates. Reducing candidate rows
lowers general matrix multiplication (GEMM) and LM-head computation and
intermediate-data traffic, but must preserve tokens that advance generation.

Our fixed verification-width experiment retains full-length drafting and
verifies the first 4, 6, 8, 10, 12, 14, or 16 slots per request.
Each prefix includes one known root token, the anchor, followed by draft candidates.
All requests use the same width without a predictor.

\begin{figure}[!t]
\centering
\includegraphics[width=0.95\columnwidth]{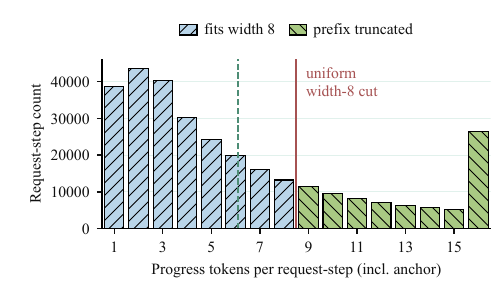}
\caption{DFlash progress on 306,215 request-steps from a mixed-dataset
training-split trace collection at $c=8$. Progress includes one anchor.
The dashed line marks the mean, 6.09. Prefixes to the right of the
width-8 boundary lose their excess tokens under uniform truncation.}
\label{fig:acceptdist}
\end{figure}

\begin{figure}[!t]
\centering
\includegraphics[width=\columnwidth]{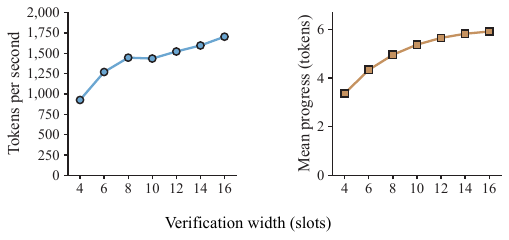}
\caption{Fixed verification widths on Qwen3-8B and GSM8K, A100, TP1,
temperature~0, $c=8$, with full-length DFlash drafting and no predictor.
Each width uses one run with 8 warmup and 256 measured requests.
Left shows output-token throughput. Right averages periodically logged
progress lengths, including the anchor. Width~16 is native DFlash.}\label{fig:naivecut}
\end{figure}

Fig.~\ref{fig:acceptdist} shows a mixed-dataset DFlash trace collection.
Per-request progress, counting one anchor plus accepted draft tokens,
has mean 6.09 and a tail reaching 16.
Fig.~\ref{fig:naivecut} shows the width comparison on GSM8K at concurrency~8.
All widths below 16 yield lower throughput than full-width verification.
Width~8 reduces throughput by 15\% and mean progress from 5.92 to 4.96 tokens.
Width~4 further lowers progress to 3.37 tokens, whereas width~14 retains
5.83 tokens but reduces selected verification slots by only 12.5\%.

\begin{figure}[!t]
\centering
\includegraphics[width=\columnwidth]{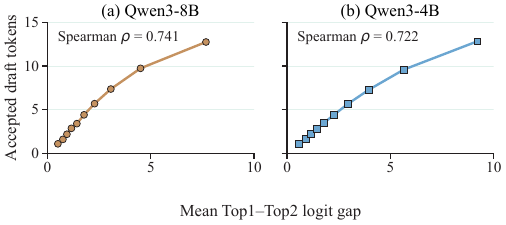}
\caption{Draft logit gaps and acceptance on 31,814 Qwen3-8B and 31,188
Qwen3-4B held-out request-steps. Each point gives mean gap and accepted draft
tokens in one of ten equal-count bins. Acceptance excludes the anchor.
Spearman correlations use all request-steps.}\label{fig:draftfeature}
\end{figure}

Although eight exceeds the mean progress, a common cutoff wastes slots on
short prefixes and truncates long ones. Actual acceptance is only known
after verification. Fully verifying blocks to choose their lengths defeats
the saving. Fig.~\ref{fig:draftfeature} shows that draft-side signals already
distinguish acceptance potential. The mean Top1--Top2 logit gap across
15 draft positions correlates positively with full-width accepted draft length,
with Spearman correlations of 0.741 on 8B and 0.722 on 4B.
These signals are available before target verification, motivating request-level
prefix prediction.

DSpark couples acceptance estimation to drafter training. Its confidence head
is jointly trained with the backbone drafter and sequential module~\cite{dspark}.
DSpark's hardware-aware prefix scheduler uses held-out sequential temperature
scaling (STS) to calibrate confidence and a steps-per-second (SPS) curve
profiled at engine initialization to select verification budgets.
To overlap scheduling with execution, its asynchronous scheduler chooses
the current total budget using confidence from two steps earlier, while
ranking current candidates by their current scores~\cite{dspark}.
Thus, current candidate ranking operates within a capacity chosen from historical
estimates, which may differ from current prefix survival.

In contrast, these signals come directly from the
existing DFlash drafter, motivating acceptance prediction independent of drafter
training. DScale trains only a separate 112K-parameter predictor on
draft logits, hidden-state features, and batch context, preserving the drafter's architecture,
weights, and full draft length. This \emph{draft-training-free} design adds
request-level budget allocation to the existing drafter without STS calibration
or SPS profiling.

\textbf{\emph{Opportunity. Predict acceptance before verification from
draft signals, independently of drafter training and without STS calibration
or SPS profiling. Retain full-length drafting and prioritize verification slots
for candidate prefixes predicted to be accepted within a preset reduced shared
budget.}}

\subsection{Adapting Dynamic Prefixes to Fixed Verification Graphs}\label{sec:motiv:c3}

Shorter logical prefixes do not automatically produce smaller graph executions.
For $N$ live requests, DFlash's native interface reserves $N\times16$ candidate slots.
Padding each shortened prefix back to its reserved width leaves GEMM and
projection dimensions unchanged. Realizing the savings from Challenge~2
therefore requires reducing the candidate-row capacity processed by the graph,
not merely marking shorter request lengths.

The observed mean progress of 6.09 motivates testing $8N$ slots instead
of $16N$, halving GEMM and projection input rows at the same request count.
The distribution's long tail argues against imposing eight slots on every request.
This calls for sharing the smaller capacity among unequal prefixes
while retaining full-length drafting.
Here, halving concerns each verification graph's candidate slots, not its
request capacity or the draft length.

Smaller capacity still leaves a layout problem.
In Challenge~3 of Fig.~\ref{fig:dflashintro}, lengths change from
$[4,12,8]$ to $[10,8,6]$. Both use 24 slots, but R2's start moves from 4 to 10.
Its old slice would mix requests, while equal-width padding expands 24 rows back to 48.
The changed boundaries must consistently identify verification inputs,
their sequence positions and KV references, and the outputs used for acceptance.

CPU re-slicing must read back GPU-computed lengths, reconstruct request
boundaries, and prepare the next verification inputs before submission.
Verification therefore waits on host preparation, placing CPU work on the GPU
critical path; if other computation cannot hide it, bubbles appear between
successive GPU stages. Compact verification must reduce candidate rows without
per-step host synchronization for slice reconstruction.

DSpark packs unequal prefixes and selects the smallest token bucket covering
their scheduled total from graphs captured at initialization~\cite{dspark}.
For example, reducing a 16-request batch from 111 to 97 verification rows
still selects the 112-slot graph. The 96-slot graph becomes usable only
when the total falls to 96 or below.
Thus, fine-grained prefix pruning produces only bucket-grained capacity reductions.
Within a bucket, fewer admitted candidates do not shrink the captured shape.
Optional bucket filling spends the spare capacity on additional candidates
instead of reducing it.

\textbf{\emph{Opportunity. Let unequal-length prefixes share halved verification
capacity while retaining full-length drafting. Keep request boundaries consistent
across verification and acceptance as lengths change, and reuse captured graphs
without per-step CPU length readback or slice reconstruction.}}

\section{System Design and Implementation}\label{sec:design}

\subsection{Overview}\label{sec:overview}

We implement DScale on top of SGLang v0.5.14.
For a fixed workload $\mathcal{R}$ and generation settings, let $\pi$ assign
verification lengths to active requests at each step.
$\mathrm{OutputTokens}_r(\pi)$ counts the output tokens returned for measured
request $r$, and $\mathrm{ElapsedTime}(\pi)$ is the measurement window's
wall-clock duration. The system objective is
\begin{equation}\label{eq:systemobjective}
\max_{\pi}\ \frac{\displaystyle\sum_{r\in\mathcal{R}}
\mathrm{OutputTokens}_r(\pi)}{\mathrm{ElapsedTime}(\pi)}.
\end{equation}
SGLang forms the active batch, and $\pi$ allocates verification slots within it.

\begin{figure*}[!t]
\centering
\includegraphics[width=0.98\textwidth]{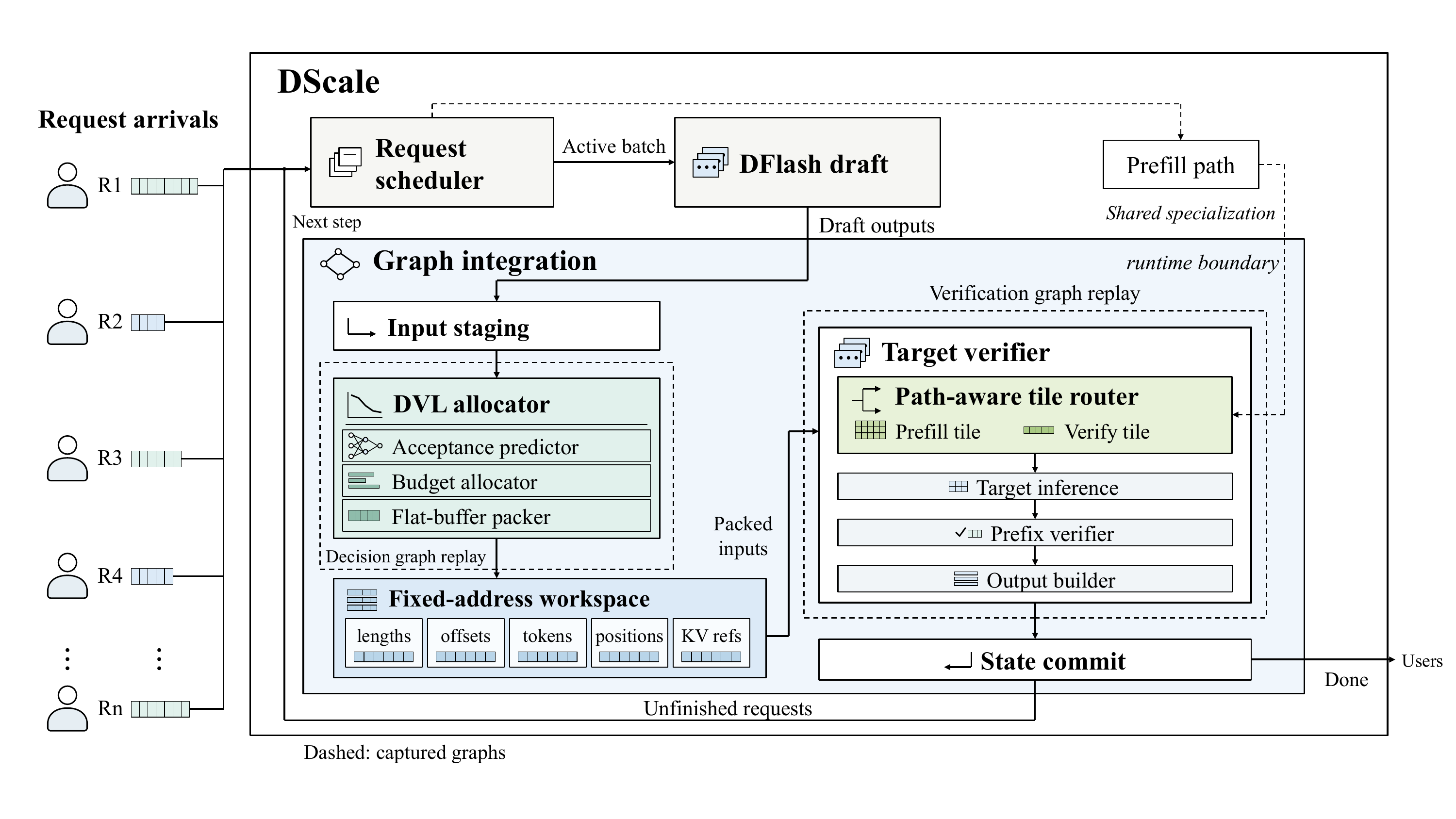}
\caption{DScale request flow and module organization.
Dashed boxes mark captured graphs, excluding input staging and state commit.
The dashed prefill branch shares tile specialization.
Lower and right paths return unfinished requests to the scheduler and
completed results to users, respectively.}\label{fig:arch}
\end{figure*}

Fig.~\ref{fig:arch} shows DScale's request flow.
The Request scheduler batches prefilled arrivals with unfinished requests.
DFlash drafter generates full-length candidate blocks, and Input staging
prepares replay inputs. Within the DVL allocator, the Acceptance predictor
estimates prefix lengths, the Budget allocator redistributes the
shared budget, and the Flat-buffer packer writes selected prefixes into
the Fixed-address workspace (Section~\ref{sec:dvl}).
Here KV refs are cache locations for this step's anchor and draft candidates.

The Target verifier consumes packed inputs through its Path-aware tile router,
retaining separate prefill specialization (Section~\ref{sec:blockm}).
Target inference evaluates candidates, the Prefix verifier determines accepted
prefixes, and the Output builder prepares output tokens.
State commit updates token and cache state, returning unfinished requests to
the scheduler and completed results to users.
Graph integration connects the two replays through the shared workspace
(Section~\ref{sec:graphintegration}).

\subsection{Path-Aware Tile Routing}\label{sec:blockm}

To address Challenge~1 in Section~\ref{sec:motiv:c1}, DScale selects tiles by path.
Maximum per-request query length $\ell_{\max}\le W=16$ selects tile height $B=16$ for verification.
Prefill and longer queries retain hardware-selected $B_{\rm hw}=128$ here for reuse.
For $N_{\rm run}$ requests and $H$ heads,
the runtime selects matching warps and $\mathrm{grid}=(N_{\rm run},H,\lceil\ell_{\max}/B\rceil)$.
Each computation block handles one request's query tile for one head.
Bucketed replay sets $N_{\rm run}=N_{\mathrm{bs}}$.

For a request with query length $\ell>0$, tile height $B$ gives the following
query-row capacity and inactive-row fraction
\begin{equation}\label{eq:tilecapacity}
\mathrm{rows}(\ell,B)=B\left\lceil\frac{\ell}{B}\right\rceil,\qquad
\mathrm{padding}(\ell,B)=1-\frac{\ell}{\mathrm{rows}(\ell,B)}.
\end{equation}
With $\ell=16$, $B=16$ fits the block. Larger tiles add inactive rows.
Smaller tiles remove no further padding and split key-value reuse across blocks.
Section~\ref{sec:ablation} evaluates this trade-off between padding and reuse.

Tile routing removes padding beyond the maximum verify width.
DVL separately packs $M=N\times8$ verification rows for GEMM and the
LM-head (Section~\ref{sec:dvl}).

\subsection{DVL Budget Allocation}\label{sec:dvl}

To address Challenge~2 in Section~\ref{sec:motiv:c2}, DVL uses draft signals
available before verification to predict acceptance and allocate unequal prefixes under a halved
shared budget, retaining full-length drafting.
For live count $N$, request-bucket capacity $N_{\mathrm{bs}}$, and block
width $W=16$ including the anchor, the captured capacity and active budget are
\begin{equation}\label{eq:verifybudget}
M_{\mathrm{bs}}=\frac{WN_{\mathrm{bs}}}{2},\qquad
M_{\rm live}=\frac{WN}{2},\qquad \sum_{i=1}^{N}k_i=M_{\rm live}.
\end{equation}
Lengths $k_i\in\{1,\ldots,W\}$ can span the full block, with sentinels at the
bucket tail. Tile selection uses $W=16$.

\textbf{Acceptance predictor.} Each request combines signals from 15 new
draft positions and batch context. Each position contributes the highest
logit, its 15 gaps to ranks 2--16, the top token's log-probability,
and distribution entropy. These $1+15+1+1=18$ preference and uncertainty
values give $15\times18=270$ confidence features in draft order.

Fixed matrices project each hidden-state vector to 64 values and each
mean-centered vocabulary-logit vector to 32.
Concatenating positions gives $15\times64=960$ hidden-state and
$15\times32=480$ logit features.
The 25-value $\mathrm{batchContext}_i$ uses live count $N$ in its first entry,
an initial-state flag of 1 in its fourth, and zeros elsewhere.
Appending it gives $270+960+480+25=1735$ features per request
\begin{equation}\label{eq:features}
\begin{aligned}
\mathrm{features}_i={}&\operatorname{Concat}(\mathrm{confidence}_i,
 \mathrm{hiddenProj}_i,\\
&\mathrm{logitProj}_i,\mathrm{batchContext}_i)\in\mathbb{R}^{1735}.
\end{aligned}
\end{equation}
Before verification, a $1735\text{--}64\text{--}15$ multilayer perceptron (MLP)
with ReLU produces one output per new draft position
\begin{equation}\label{eq:predictor}
\mathrm{predLogits}_i=\operatorname{Predictor}(\mathrm{features}_i)
\in\mathbb{R}^{W-1}.
\end{equation}

% EARLY EXPERIMENT FLOATS alg:dvl
\begin{algorithm}[!ht]
\caption{DVL allocation and flat-pack}\label{alg:dvl}
\begin{algorithmic}[1]
\REQUIRE $\mathrm{features},\ \mathrm{tokens},\ \mathrm{positions},\ \text{KV refs}$
\STATE \hspace{\algorithmicindent}$1\le N\le N_{\mathrm{bs}},\ W=16,\ M_{\rm live}=8N,\ M_{\rm bs}=8N_{\mathrm{bs}}$
\ENSURE $\mathrm{buffer}_{\mathrm{bs}};\quad\sum_{i=1}^{N}k_i=M_{\rm live}$
\STATE $\mathrm{predLogits}\gets\operatorname{Predictor}(\mathrm{features})$\label{ln:dvl:scores}
\STATE $Q\gets[\mathbf{1},\operatorname{CumMin}_{j}(\sigma(\mathrm{predLogits}))]$
\STATE $k_i^{(0)}\gets\min\{W,\max\{1,\lfloor\sum_{j=1}^{W-1}q_{i,j}+\tfrac12\rfloor\}\}$\label{ln:dvl:initial}
\STATE $\mathbf{k}_{1:N}\gets\mathbf{k}^{(0)}$; $\Delta\gets M_{\rm live}-\sum_{i=1}^{N}k_i$\label{ln:dvl:deficit}
\WHILE{$\Delta>0$}\label{ln:dvl:grow}
  \STATE $i^\ast\gets\arg\max_{1\le i\le N,\ k_i<W}q_{i,k_i}$
  \STATE $(k_{i^\ast},\Delta)\gets(k_{i^\ast}+1,\Delta-1)$
\ENDWHILE\label{ln:dvl:growend}
\WHILE{$\Delta<0$}\label{ln:dvl:shrink}
  \STATE $i^\ast\gets\arg\min_{1\le i\le N,\ k_i>1}q_{i,k_i-1}$
  \STATE $(k_{i^\ast},\Delta)\gets(k_{i^\ast}-1,\Delta+1)$
\ENDWHILE\label{ln:dvl:adjustend}
\STATE $\mathrm{offset}_{1:N+1}\gets(0,\operatorname{cumsum}(\mathbf{k}_{1:N}))$\label{ln:dvl:offset}
\FORALL{$(i,j,\mathrm{field}):\ 1\le i\le N,\ 0\le j<k_i$}\label{ln:dvl:packstart}
  \STATE $s\gets\mathrm{offset}_i+j$
  \STATE $\mathrm{buffer}_{\mathrm{bs}}.\mathrm{field}[s]\gets\mathrm{field}_i[j]$
\ENDFOR\label{ln:dvl:packend}
\STATE $(k_i,\mathrm{offset}_{i+1})\gets(0,M_{\rm live}),\quad \forall\,N<i\le N_{\mathrm{bs}}$\label{ln:dvl:sentinel}
\STATE $\mathrm{buffer}_{\mathrm{bs}}[s]\gets(0,0,\text{KV refs}_1[0]),\quad \forall\,M_{\rm live}\le s<M_{\mathrm{bs}}$\label{ln:dvl:sentinelend}
\STATE \textbf{return} $\mathrm{buffer}_{\mathrm{bs}}$\label{ln:dvl:return}
\end{algorithmic}
\end{algorithm}

\begin{figure}[!t]
\centering
% Native source: DVL_allocator_score_interface_v4.pptx (2026-09-14).
\includegraphics[width=\columnwidth]{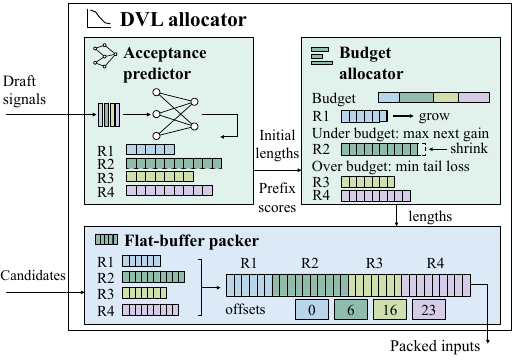}
\caption{DVL allocator. The predictor supplies initial lengths and prefix scores
to the Budget allocator. The Flat-buffer packer stores prefixes and offsets.
Short strips select from full draft blocks. Grow and shrink are alternative
budget conditions. Colors identify requests.}\label{fig:dvl}
\end{figure}

Sigmoid $\sigma$ gives acceptance scores, with $q_0=1$ for the known anchor.
Acceptance requires accepted predecessors. Cumulative minima therefore
enforce nonincreasing prefix scores
\begin{equation}\label{eq:surv}
\begin{aligned}
q_0 &= 1,\\
q_j &= \min\!\bigl(q_{j-1},\;\sigma(\mathrm{predLogits}_j)\bigr),\quad j=1,\dots,W{-}1 .
\end{aligned}
\end{equation}
Summing request~$i$'s 15 prefix scores estimates accepted draft count $\hat\ell_i$.
Rounding and clipping to $[1,W]$ initializes heuristic slot count $k_i^{(0)}$,
including the anchor
\begin{equation}\label{eq:elen}
\begin{aligned}
\hat\ell_i &= \sum_{j=1}^{W-1}q_{i,j},\\
k_i^{(0)} &= \min\!\left\{W,\max\!\left\{1,
  \left\lfloor\hat\ell_i+\tfrac12\right\rfloor\right\}\right\}.
\end{aligned}
\end{equation}
Thus $\hat\ell_i=6.2$ initializes one anchor and five new candidates.
The Budget allocator adjusts this seed using \(Q=(q_{i,j})\)
and the shared budget in Fig.~\ref{fig:dvl}.

Position $j$ is labeled 1 if full-width verification accepts at least $j$
new tokens. Training minimizes mean binary cross-entropy plus 0.02 times
mean squared error between the raw sigmoid sum and accepted draft count,
before monotonicity enforcement. The supplement gives reproduction details.

\textbf{Budget allocator.} The prefix-score objective for the scheduler's
$N$ requests is
\begin{equation}\label{eq:marginal}
\max_{\{k_i\}} \;\sum_{i=1}^{N}\sum_{j=0}^{k_i-1} q_{i,j}
\quad\text{s.t.}\quad \sum_{i=1}^{N} k_i = M_{\rm live},\;\; 1 \le k_i \le W .
\end{equation}
DVL starts at \(k_i=k_i^{(0)}\) with
$\Delta=M_{\rm live}-\sum_i k_i$. Next-slot gain \(q_{i,k_i}\) and
last-slot loss \(q_{i,k_i-1}\) determine the request to adjust
\begin{equation}\label{eq:budgetchoice}
i^\ast=
\begin{cases}
\displaystyle\arg\max_{i:\,k_i<W}q_{i,k_i},&\Delta>0,\\
\displaystyle\arg\min_{i:\,k_i>1}q_{i,k_i-1},&\Delta<0.
\end{cases}
\end{equation}
The adjustment loop executes in one GPU computation block within the decision graph,
without host round trips or inter-block synchronization.

\begin{proposition}\label{prop:budgetadjustment}
For integer seeds $1\le k_i^{(0)}\le W$, integer budget
$N\le M_{\rm live}\le NW$, nonincreasing scores, and
$\Delta_0=M_{\rm live}-\sum_i k_i^{(0)}$,
Algorithm~\ref{alg:dvl} preserves prefixes and anchors and meets the budget
in $|\Delta_0|$ updates. It maximizes Eq.~\eqref{eq:marginal} over integer lengths subject additionally
to $k_i\ge k_i^{(0)}$ for every $i$ if $\Delta_0>0$, or
$k_i\le k_i^{(0)}$ if $\Delta_0<0$. Zero gap retains the seed.
\end{proposition}
The proof is provided in the supplementary material.

Algorithm~\ref{alg:dvl}'s $\mathbf{k}$ and $\mathrm{offset}$ reside in
$\mathrm{buffer}_{\mathrm{bs}}$, Fig.~\ref{fig:arch}'s Fixed-address workspace.
$\text{KV refs}$ identify candidate cache locations.
Lines~\ref{ln:dvl:scores}--\ref{ln:dvl:initial} predict logits, apply sigmoid and cumulative
minimum, then round draft-only score sums into initial lengths.
Line~\ref{ln:dvl:deficit} sets the budget gap $\Delta$.
Lines~\ref{ln:dvl:grow}--\ref{ln:dvl:growend} grow the highest-scoring next slot.
Lines~\ref{ln:dvl:shrink}--\ref{ln:dvl:adjustend} remove the lowest-scoring tail.
Both preserve anchored prefixes and stop at $\Delta=0$.

Line~\ref{ln:dvl:offset} computes offsets. Lines~\ref{ln:dvl:packstart}--\ref{ln:dvl:packend} pack token IDs, positions, and cache
references with identical row mappings.
\textbf{For all} permits independent parallel GPU iterations.
Lines~\ref{ln:dvl:sentinel}--\ref{ln:dvl:sentinelend} zero inactive lengths, set their offsets to $M_{\rm live}$,
and initialize tail placeholders. Line~\ref{ln:dvl:return} returns the buffer with $8N$ live slots.

\textbf{Flat-buffer packer.} Allocated prefixes of lengths \(k_i\) occupy
$\mathrm{buffer}_{\mathrm{bs}}$ of capacity $M_{\mathrm{bs}}$, using prefix-sum offsets
\begin{equation}\label{eq:offset}
\mathrm{offset}_i \;=\; \sum_{j<i} k_j
\end{equation}
The packer applies this mapping to each $\mathrm{field}$ in tokens, positions,
and cache references
\begin{equation}\label{eq:packmap}
\mathrm{buffer}_{\mathrm{bs}}.\mathrm{field}[\mathrm{offset}_i+j]
=\mathrm{field}_i[j],\quad 0\le j<k_i.
\end{equation}
Each attention block handles one request and head, masking rows beyond $k_i$.
Acceptance reuses the offsets
(Section~\ref{sec:graphintegration}).
In Fig.~\ref{fig:dvl}, Initial lengths $k^{(0)}=[5,10,7,9]$ total 31 against
budget 32. Prefix scores denote $Q$. Growing R1 gives adjusted lengths
$k=[6,10,7,9]$, offsets $[0,6,16,23]$, and end boundary 32.
R2's dashed tail illustrates alternative shrinkage.

\textbf{Complexity analysis.} Given draft features, prediction costs
$O(N_{\mathrm{bs}}dh)$ for input dimension $d=1735$ and hidden dimension $h=64$.
Rounding costs $O(N_{\mathrm{bs}}W)$.
Packing scans request boundaries for each slot, costing $O(M_{\mathrm{bs}}N_{\mathrm{bs}})$,
including $O(M_{\mathrm{bs}})$ row writes and tail initialization.
At most $O(NW)$ adjustments each scan $O(N_{\mathrm{bs}}W)$ scores, yielding
\begin{equation}\label{eq:allocatorwork}
\mathrm{Work}_{\rm DVL}=O(N_{\mathrm{bs}}dh+NN_{\mathrm{bs}}W^2+M_{\mathrm{bs}}N_{\mathrm{bs}}).
\end{equation}
For $W=16$, the adjustment count is $|8N-\sum_i k_i^{(0)}|\le7N$.
Each batch-score scan costs $O(N_{\mathrm{bs}})$, giving
$O(NN_{\mathrm{bs}})$ adjustment work.
Equation~\eqref{eq:allocatorwork} bounds aggregate GPU-thread work,
not parallel elapsed time.
On GSM8K at $c=8$, combined GPU execution of the Acceptance predictor,
Budget allocator, and Flat-buffer packer accounts for 3.08\% of DScale's
complete decode-step time on Qwen3-8B and 4.08\% on Qwen3-4B, as measured in
Fig.~\ref{fig:allocatorkernelcosts}(c).

\subsection{Graph Integration for Ragged Decisions}\label{sec:graphintegration}

To address Challenge~3 in Section~\ref{sec:motiv:c3}, a fixed-address workspace
carries changing boundaries through packing, verification, and acceptance.
Each workspace tensor $\mathrm{array}_{\mathrm{bs}}$ retains its address and shape across steps
\begin{equation}\label{eq:graphinvariant}
\begin{aligned}
\operatorname{addr}(\mathrm{array}_{\mathrm{bs}}^{(t+1)})
 &=\operatorname{addr}(\mathrm{array}_{\mathrm{bs}}^{(t)}),\\
\operatorname{shape}(\mathrm{array}_{\mathrm{bs}}^{(t+1)})
 &=\operatorname{shape}(\mathrm{array}_{\mathrm{bs}}^{(t)}).
\end{aligned}
\end{equation}
After full-length drafting, Input staging selects from finite captured capacities $\mathcal B$
\begin{equation}\label{eq:bucketselect}
N_{\mathrm{bs}}=\min\{b\in\mathcal{B}\mid b\ge N\}.
\end{equation}
Inactive lengths are zero. Including tail slots excluded from acceptance,
capacity remains half that of DFlash's same request bucket
\begin{equation}\label{eq:bucketpadding}
M_{\rm live}+8(N_{\mathrm{bs}}-N)=8N_{\mathrm{bs}}
=\frac{16N_{\mathrm{bs}}}{2}.
\end{equation}

For greedy decoding, Algorithm~\ref{alg:graph}, lines~\ref{ln:graph:init}--\ref{ln:graph:captureend},
allocates workspaces and captures Fig.~\ref{fig:arch}'s two pipelines.
Features constructs predictor inputs, and DVL invokes Algorithm~\ref{alg:dvl}.
Target, Accept, and Output denote tile-routed Target inference,
the Prefix verifier, and the Output builder.
Lines~\ref{ln:graph:stage}--\ref{ln:graph:decision} perform Input staging and decision replay.
The request state supplies context lengths $\mathrm{ctxLen}$,
request IDs $\mathrm{reqID}$, and cache mappings $\mathrm{cacheMap}$.
Line~\ref{ln:graph:order} orders GPU reads after decision and prior cache writes.

% EARLY EXPERIMENT FLOATS alg:graph
\begin{algorithm}[!ht]
\caption{Fixed-address graph capture and replay}\label{alg:graph}
\begin{algorithmic}[1]
\REQUIRE $\mathcal B,\ \mathrm{requests},\ \mathrm{draft},\ \mathrm{requestState}$
\STATE $\emptyset\ne\mathcal B\subset\mathbb Z_{>0},\quad 1\le|\mathrm{requests}|\le\max\mathcal B$
\ENSURE Committed outputs and remaining requests
\STATE \textit{Capture once}
\FOR{$N_{\mathrm{bs}}\in\mathcal B$}\label{ln:graph:init}
\STATE $\mathrm{buffer}_{\mathrm{bs}}\gets\operatorname{Alloc}(N_{\mathrm{bs}},8N_{\mathrm{bs}})$
\STATE $\mathcal G^{\rm dec}_{\mathrm{bs}}\gets
\operatorname{Capture}_{\mathrm{bs}}[\mathrm{Features}\rightarrow\mathrm{DVL}]$
\STATE $\mathcal G^{\rm verify}_{\mathrm{bs}}\gets
\operatorname{Capture}_{\mathrm{bs}}[\mathrm{Target}\rightarrow\mathrm{Accept}\rightarrow\mathrm{Output}]$
\ENDFOR\label{ln:graph:captureend}
\STATE \textit{Replay each step}
\STATE $N\gets|\mathrm{requests}|,\quad N_{\mathrm{bs}}\gets\min\{b\in\mathcal B\mid b\ge N\}$\label{ln:graph:stage}
\STATE $\mathrm{buffer}_{\mathrm{bs}}.\mathrm{inputs}\gets(N,\mathrm{draft},\mathrm{requestState})$
\STATE $\operatorname{Replay}(\mathcal G^{\rm dec}_{\mathrm{bs}})$\label{ln:graph:decision}
\STATE $\operatorname{WaitGPU}(\mathcal G^{\rm dec}_{\mathrm{bs}},\mathrm{KVWrite}_{\rm prior})$\label{ln:graph:order}
\FORALL{$1\le i\le N_{\mathrm{bs}}$}\label{ln:graph:metadata}
\STATE $(\mathrm{qStart}_i,\mathrm{qEnd}_i)\gets
(\mathrm{offset}_i,\mathrm{offset}_i+k_i)$
\ENDFOR\label{ln:graph:queryend}
\FORALL{$1\le i\le N,\ 0\le j<\mathrm{ctxLen}_i$}\label{ln:graph:cachebegin}
\STATE $\mathrm{historyKV}_i[j]\gets\mathrm{cacheMap}[\mathrm{reqID}_i,j]$
\ENDFOR\label{ln:graph:metadataend}
\STATE $\operatorname{Replay}(\mathcal G^{\rm verify}_{\mathrm{bs}})$\label{ln:graph:verify}
\STATE $\mathrm{outputs}\gets\operatorname{ReadOutputs}(\mathrm{buffer}_{\mathrm{bs}})$\label{ln:graph:commit}
\FORALL{$\mathrm{req}_i\in\mathrm{requests}$}\label{ln:graph:statebegin}
\STATE $\operatorname{Append}(\mathrm{req}_i,\mathrm{outputs}_i.\mathrm{tokens})$
\STATE $\mathrm{req}_i.\mathrm{length}\gets\mathrm{outputs}_i.\mathrm{length}$
\STATE $\mathrm{req}_i.\mathrm{next}\gets\mathrm{outputs}_i.\mathrm{next}$
\ENDFOR\label{ln:graph:stateend}
\STATE $\mathrm{requests}\gets\{\mathrm{req}_i\in\mathrm{requests}\mid
\neg\operatorname{Finished}(\mathrm{req}_i)\}$\label{ln:graph:filter}
\STATE \textbf{return} $\mathrm{requests},\ \mathrm{outputs}$\label{ln:graph:return}
\end{algorithmic}
\end{algorithm}

Lines~\ref{ln:graph:metadata}--\ref{ln:graph:queryend} set $\mathrm{qStart}$ and $\mathrm{qEnd}$ from $k$ and $\mathrm{offset}$.
Lines~\ref{ln:graph:cachebegin}--\ref{ln:graph:metadataend} gather $\mathrm{ctxLen}_i$ committed-context indices from
$\mathrm{cacheMap}$ row $\mathrm{reqID}_i$ into $\mathrm{historyKV}_i$.
These views reside in persistent $\mathrm{buffer}_{\mathrm{bs}}$ metadata.
\textbf{For all} executes independent operations in parallel on the GPU.
$\mathrm{historyKV}$ indexes unmoved context KV, while candidate $\text{KV refs}$ identify new writes.
Line~\ref{ln:graph:verify} replays verification, acceptance, and output construction.
Attention retains its grid and masks rows by
\begin{equation}\label{eq:rowvalid}
\mathrm{valid}_i(j)=\mathbf{1}\{1\le i\le N\ \land\ 0\le j<k_i\}.
\end{equation}
Valid rows read $\mathrm{offset}_i+j$ with request-local causal masking
and the supplied cache references.

\textbf{Prefix verifier.} Let $P$ denote the target decoding policy's
next-token distribution conditioned on the current prefix.
A deterministic top-1 draft candidate $x$ is accepted with probability $P(x)$.
At the first rejection, draw a successor from $P$ with $x$ removed and
the remaining probabilities renormalized.
If all $k_i-1$ selected candidates pass, use the unmodified target distribution.
After $a_i$ accepted candidates, row $\mathrm{offset}_i+a_i$ supplies the successor.
Output the accepted candidates followed by this successor, which anchors the next step.
DVL chooses lengths independently of acceptance random numbers.
Section~3 of the supplementary material derives the resulting output distribution.

The Output builder also materializes projected draft cache.
Line~\ref{ln:graph:commit} retrieves tokens, committed lengths, and successors.
Lines~\ref{ln:graph:statebegin}--\ref{ln:graph:stateend} implement State commit outside capture,
appending valid tokens under stopping rules and updating committed lengths.
Lines~\ref{ln:graph:filter}--\ref{ln:graph:return} remove finished requests and return remaining requests and outputs.

% EARLY EXPERIMENT FLOATS fig:mainthru
\begin{figure*}[!t]
\centering
\includegraphics[width=0.98\textwidth]{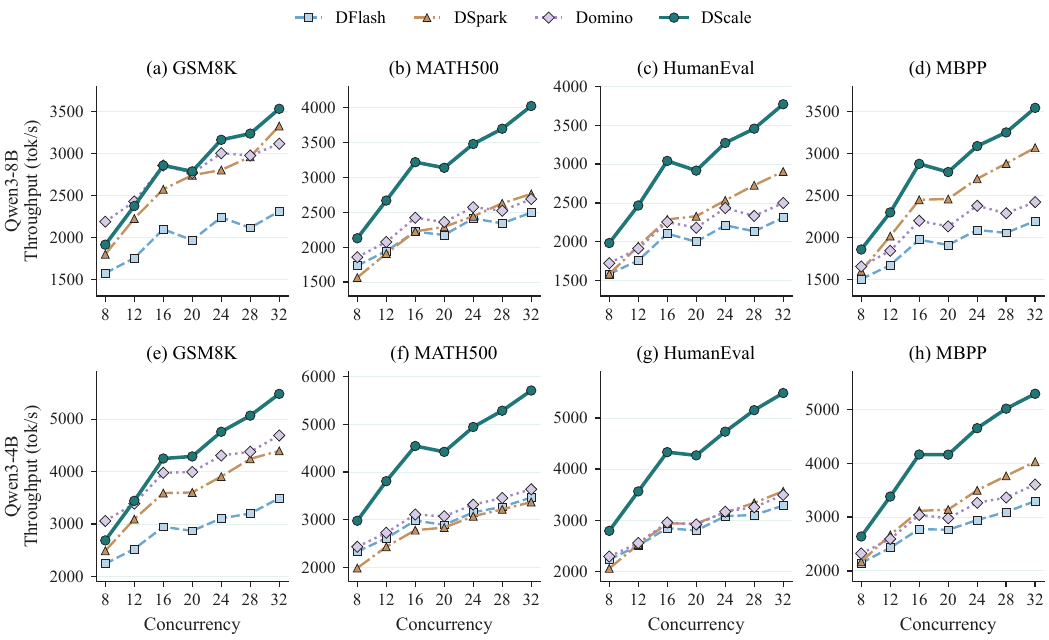}
\caption{Token throughput at concurrency 8--32. Top and bottom rows show
Qwen3-8B and Qwen3-4B. Columns show GSM8K, MATH500, HumanEval, and MBPP.}
\label{fig:mainthru}
\end{figure*}

% EARLY EXPERIMENT FLOATS fig:lat
\begin{figure*}[!t]
\centering
\includegraphics[width=0.95\textwidth]{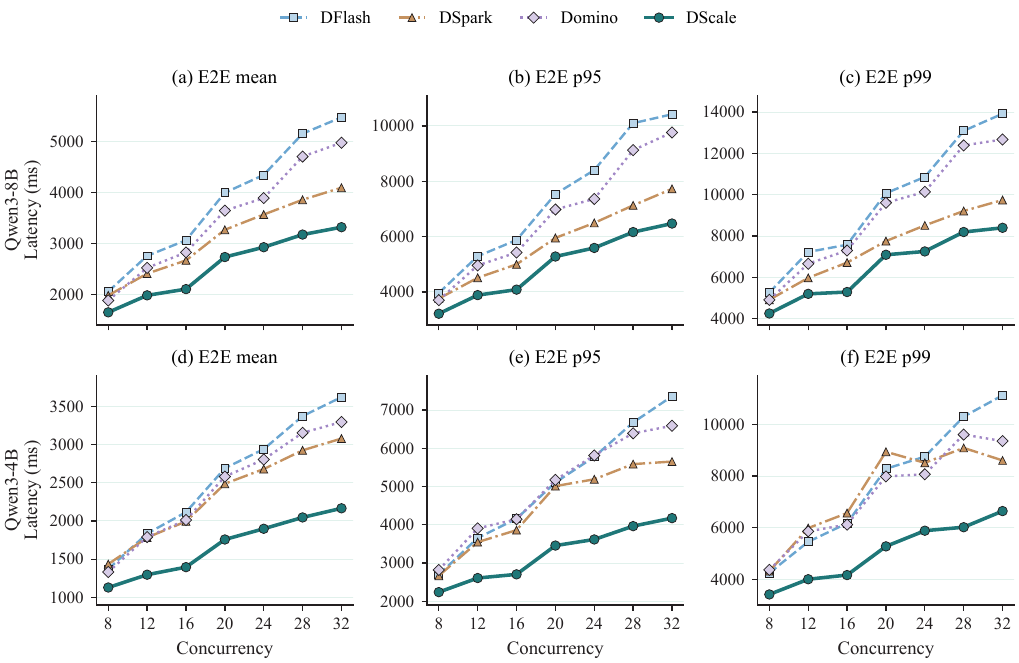}
\caption{End-to-end request latency at concurrency 8--32. Top and bottom rows
show Qwen3-8B and Qwen3-4B. Columns show mean, p95, and p99.
Each point is the median across four datasets of their five-repetition medians.}\label{fig:lat}
\end{figure*}

% EARLY EXPERIMENT FLOATS fig:ablation
\begin{figure}[!t]
\centering
\includegraphics[width=0.98\columnwidth]{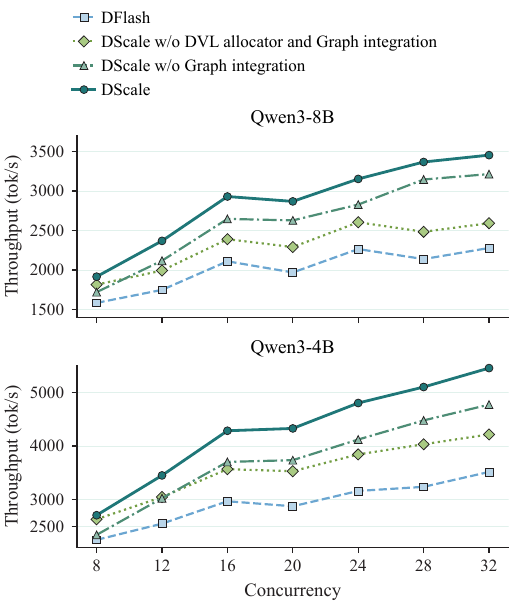}
\caption{GSM8K cumulative ablation using five-repetition medians,
successively adding tile routing, DVL, and Graph integration.}\label{fig:ablation}
\end{figure}

% EARLY EXPERIMENT FLOATS fig:allocatorretention
\begin{figure}[!t]
\centering
\includegraphics[width=0.98\columnwidth]{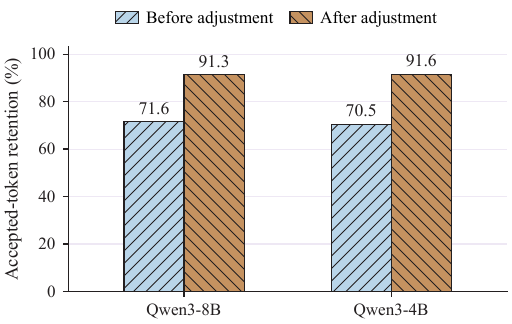}
\caption{Accepted-token retention before and after budget adjustment on
31,814 validation request-steps for Qwen3-8B and 31,188 for Qwen3-4B.
Features and full-width labels are fixed. Each collection-step cohort of $N$
retained rows receives $8N$ slots after adjustment.}
\label{fig:allocatorretention}
\end{figure}

\section{Experimental Evaluation and Discussion}\label{sec:eval}

\subsection{Experimental Setup}\label{sec:setup}

\textbf{Hardware and models.} The main-paper benchmarks run on a
single A100-40GB GPU with tensor parallelism~1 and generation temperature~0.
We evaluate Qwen3-8B and Qwen3-4B~\cite{qwen2025qwen3} as target models.

\textbf{Workloads.} We use four math and code benchmarks.
\begin{itemize}
\item GSM8K~\cite{cobbe2021training} covers multi-step grade-school math.
\item MATH500~\cite{lightman2024math500} has 500 competition math problems.
\item HumanEval~\cite{chen2021evaluating} tests docstring-based completion.
\item MBPP~\cite{austin2021program} tests text-to-Python programming.
\end{itemize}

Each formal configuration runs 32 warmup requests followed by 512 measured requests,
using fixed request manifests with deterministic prompt repetitions shared across systems.
The measured manifests contain 256, 256, 164, and 255 distinct prompts for
GSM8K, MATH500, HumanEval, and MBPP, respectively.
Thinking is disabled, and the output cap is 2048 tokens.
Token throughput is the total output tokens generated by the measured requests
divided by their measurement window's wall-clock duration.
The separate motivation collections use the protocols specified with their figures.

\textbf{Concurrency sweep.} With matched seeds and prompts, we sweep
$c\in\{8,12,16,20,24,28,32\}$, report five-repetition medians, and compute
aggregate throughput ratios as geometric means over dataset--concurrency cells.
Here $c$ is both the number of closed-loop client workers and the server's
maximum running requests; each worker submits its next request upon completion
until queue exhaustion, then outstanding requests drain, while the active decode
batch can vary with scheduling.

\textbf{Baselines.} We compare four SGLang-based systems under the same request protocol.
(i)~\textbf{DFlash}~\cite{chen2026dflash} is the SGLang v0.5.14 implementation before our modifications,
using block diffusion at width~16 with the default 128-row verify tile.
(ii)~\textbf{DSpark}~\cite{dspark} is the implementation in SGLang v0.5.16,
configured with its corresponding draft model for each target.
We use v0.5.16 because v0.5.14 does not include a DSpark implementation.
(iii)~\textbf{Domino}~\cite{domino} uses its official draft configuration.
(iv)~\textbf{DScale} is our runtime built on
the SGLang v0.5.14 DFlash implementation.
DFlash and DScale use the same released DFlash draft checkpoint for each
target. DSpark and Domino use their respective target-matched draft checkpoints.
Tokenization, prompt formatting, stopping criteria,
maximum output length, and generation temperature are held fixed.

% EARLY EXPERIMENT FLOATS fig:runtimeprofile
\begin{figure*}[!t]
\centering
\includegraphics[width=0.98\textwidth]{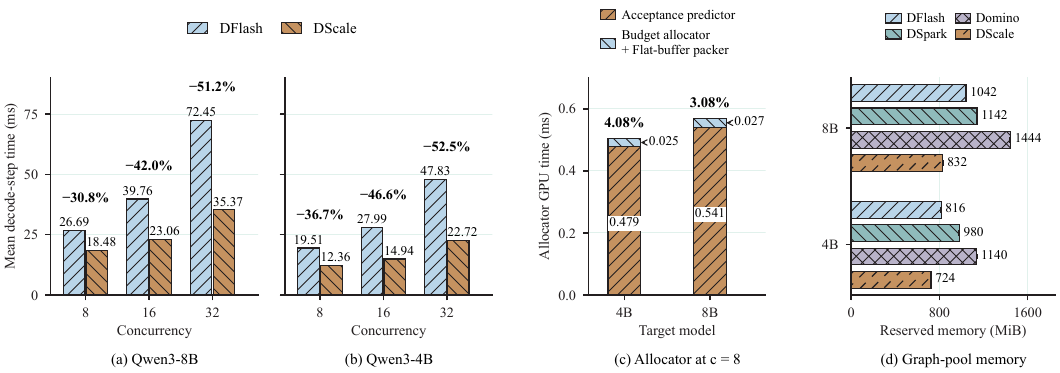}
\caption{Decode-step time, allocator costs, and graph-pool memory under TP1.
Panels a--c use GSM8K at temperature 0. Panels a and b show mean step times
and reductions from DFlash. Panel c reports $c=8$ allocator time in milliseconds
and as a percentage of DScale step time. Overlapping component intervals count
once. The Budget allocator and Flat-buffer packer are timed together.
Panel d reports reserved memory in MiB, showing means and sample standard
deviations over three independent process starts.}\label{fig:runtimeprofile}
\label{fig:allocatorkernelcosts}
\label{fig:graphpool}
\end{figure*}

\subsection{Throughput Results}

\textbf{Overall gains.}
For the results in Fig.~\ref{fig:mainthru}, taking the geometric mean of per-cell throughput ratios across all four datasets
and seven concurrencies, DScale improves over DFlash, DSpark, and Domino
by 43.9\%, 22.2\%, and 24.4\%, respectively, on Qwen3-8B,
and by 48.8\%, 37.7\%, and 32.0\% on Qwen3-4B.

\textbf{Concurrency scaling.}
Four-dataset geometric-mean gains over DFlash increase with concurrency
on both models. At $c=8$, gains are $23.2\%$ on Qwen3-8B and $24.0\%$ on Qwen3-4B.
They reach $42.8\%$ and $49.7\%$ at $c=16$, then $59.6\%$ and $62.3\%$ at $c=32$.
Against DSpark, gains at these three concurrencies are $20.4\%$, $25.8\%$,
and $23.4\%$ on Qwen3-8B, and $27.9\%$, $39.8\%$, and $43.8\%$ on Qwen3-4B.
Gains over Domino also grow from $6.7\%$ to $39.1\%$ on 8B
and from $10.5\%$ to $43.4\%$ on 4B between $c=8$ and $c=32$.

\textbf{Dataset differences.}
For each dataset, we take the geometric mean of paired throughput ratios
over all seven concurrencies from 8 to 32.
In GSM8K, MATH500, HumanEval, and MBPP order, gains over DFlash are
40.0\%, 44.1\%, 46.5\%, and 45.3\% on Qwen3-8B,
and 44.9\%, 51.4\%, 50.7\%, and 48.5\% on Qwen3-4B.
In the same dataset order, gains over DSpark are
7.7\%, 40.8\%, 28.0\%, and 14.8\% on Qwen3-8B,
and 17.4\%, 60.1\%, 46.7\%, and 30.3\% on Qwen3-4B.
Gains over Domino are 1.6\%, 33.9\%, 34.7\%, and 30.6\% on Qwen3-8B,
and 6.3\%, 44.1\%, 45.1\%, and 36.6\% on Qwen3-4B.

\textbf{Sampling and tensor parallelism.}
Supplementary Figs.~5--7 show geometric-mean gains over DFlash of 45.2\%
on Qwen3-8B and 45.8\% on Qwen3-4B under TP1 at temperature~1,
and 43.9--55.8\% under TP2 at temperatures~0 and~1.
The same predictors thus support sampling and tensor parallelism without retraining.

\subsection{End-to-End Latency}

End-to-end (E2E) request latency is measured in milliseconds.
For mean, p95, and p99, reductions are one minus the geometric mean of
DScale-to-baseline ratios of five-repetition medians across four datasets and
seven concurrencies.

Mean E2E latency reductions relative to
DFlash, DSpark, and Domino are 31.4\%, 18.4\%, and 20.2\% on Qwen3-8B,
and 33.1\%, 27.1\%, and 24.7\% on Qwen3-4B.

The p95 and p99 reductions relative to DFlash
are 30.7\% and 29.3\% on Qwen3-8B, and 32.9\% and 31.9\% on Qwen3-4B.
Relative to DSpark, they are 16.8\% and 13.1\% on Qwen3-8B,
and 26.4\% and 26.8\% on Qwen3-4B.
Relative to Domino, they are 22.3\% and 21.4\% on Qwen3-8B,
and 28.5\% and 27.0\% on Qwen3-4B.

Fig.~\ref{fig:lat} shows lower cross-dataset medians for both models than
all three baselines at every concurrency for mean, p95, and p99.
Supplementary Figs.~1--4 separate the datasets.
Geometric-mean p99 reductions over seven concurrencies range from
26.1\% to 34.6\% relative to DFlash, covering both math and code workloads.

\subsection{Component Ablation}\label{sec:ablation}

The GSM8K ablation holds the runtime, target, drafter, and request set fixed,
using temperature~0 and memory fraction~0.85.
Its DFlash control disables all three optimizations and is measured separately
from the main-comparison baseline.
Fig.~\ref{fig:ablation} adds tile routing, DVL allocation, and Graph integration
successively, using five-repetition medians, 32 warmup requests, and 512 measured
requests. Optimized variants share the model-specific predictor and projections.

Taking geometric means across the seven concurrencies, tile routing improves throughput over the
DFlash control by $14.7\%$ on Qwen3-8B and $20.8\%$ on Qwen3-4B.
Adding DVL to the tile-routed variant yields a further $11.8\%$ and
$3.9\%$, respectively. Adding Graph integration to that combined variant
provides another $9.8\%$ and $15.2\%$.

Relative to the same DFlash control, the cumulative geometric-mean gains
reach $28.2\%$ on 8B and $25.5\%$ on 4B after adding DVL, then rise to
$40.8\%$ and $44.6\%$ with all three mechanisms.
DVL's gains are more pronounced at higher concurrency.
The complete DScale system exceeds the DFlash control at every evaluated
concurrency on both models.

\textbf{Predictor quality and budget adjustment.}
Accepted-length regression on the prompt-disjoint holdout reaches
$R^2\approx0.70$ on Qwen3-8B and $0.68$ on Qwen3-4B.
We compare accepted-token retention before
and after budget adjustment on the deployed predictors' frozen validation splits,
holding logged features and histories fixed. Full-width acceptance labels
are $y_i$. For a window $k_i$ including one anchor slot, retention is
$\sum_i\min(y_i,k_i-1)/\sum_i y_i$.
As shown in Fig.~\ref{fig:allocatorretention}, raw predictor windows retain
71.6\% of accepted tokens on Qwen3-8B and
70.5\% on Qwen3-4B. Score-guided adjustment raises these to 91.3\% and 91.6\%,
respectively. For each model, the same model-specific predictor is used
before and after budget adjustment.
This single-step evaluation includes both score-guided redistribution and
budget expansion. Mean windows increase from 5.11 slots on 8B and 5.08 on 4B
to eight on both models, with a total budget of $8N$ per retained-row cohort.

% Declared ahead of Related Work so the two-column float stays near that section.

% EARLY EXPERIMENT FLOATS tab:related-capabilities
\begin{table*}[t]
\centering
\caption{Verification-side serving capabilities of the compared execution paths.}
\label{tab:related-capabilities}
\footnotesize
\begin{tabular*}{\textwidth}{@{\extracolsep{\fill}}lcccccc@{}}
\toprule
Method & \shortstack{Drafter\\reused\textsuperscript{a}}
& \shortstack{Per-request\\budget\textsuperscript{b}}
& \shortstack{Compact\\variable-length\\verify}
& \shortstack{CUDA Graph\\integration\textsuperscript{c}}
& \shortstack{Extra\\predictor\textsuperscript{d}}
& \shortstack{No hardware\\performance-curve profiling\\for budget scheduling\textsuperscript{e}} \\
\midrule
DFlash~\cite{chen2026dflash,sglang}
& $\checkmark$ & $\times$ & $\times$ & $\checkmark$ & $\times$ & $\checkmark$ \\
DSpark~\cite{dspark}
& $\times$ & $\checkmark$ & $\checkmark$ & $\checkmark$ & $\checkmark$ & $\times$ \\
Domino~\cite{domino}
& $\times$ & $\times$ & $\times$ & $\checkmark$ & $\times$ & $\checkmark$ \\
\textbf{DScale}
& $\checkmark$ & $\checkmark$ & $\checkmark$ & $\checkmark$ & $\checkmark$ & $\checkmark$ \\
\bottomrule
\end{tabular*}
\par\smallskip
\begin{minipage}{\textwidth}
$\checkmark$ means the property holds, and $\times$ means it does not.
The baselines follow the experimental setup in Section~\ref{sec:setup}.
\textsuperscript{a}Drafter reuse means retaining the existing DFlash drafter's architecture and
weights. \textsuperscript{b}Per-request budget denotes adaptive prefix-length selection before verification.
\textsuperscript{c}Graph integration indicates a captured execution path.
\textsuperscript{d}Extra predictor denotes an acceptance estimator for budget selection.
DSpark integrates a confidence head trained with its drafter, whereas DScale
trains a separate predictor and retains the drafter.
\textsuperscript{e}DSpark uses a premeasured SPS curve for budget scheduling.
\end{minipage}
\end{table*}

\textbf{Current-runtime decode profile.}
We profile DFlash and DScale CUDA workers on GSM8K with Nsight Systems,
using the TP1 protocol in Section~\ref{sec:setup}, memory fraction~0.85,
and one run per $c\in\{8,16,32\}$.
We average 700 step intervals after discarding 40.
Intervals span consecutive nonempty decode-worker entries, including host
gaps but excluding prefill and idle boundaries.
Kernels are timed in full even when they outlast the CPU step.
Fig.~\ref{fig:runtimeprofile}(a) and (b) support the throughput gains with
30.8--52.5\% shorter intervals, using $1-T_{\mathrm{DScale}}/T_{\mathrm{DFlash}}$.
Under TP2 at $c=32$, Supplementary Fig.~8 shows 50.7\% and 52.8\%
reductions on Qwen3-8B and Qwen3-4B, supporting the two-GPU throughput gains.

\textbf{Allocator kernel costs.}
Fig.~\ref{fig:allocatorkernelcosts}(c) breaks down the $c=8$ costs.
The Acceptance predictor, Budget allocator, and Flat-buffer packer
together account for 4.08\% of step time on 4B and 3.08\% on 8B.
Predictor timing includes feature extraction, shared Top1 selection, and input
preparation, excluding the LM-head. The fused cost covers budget adjustment,
metadata, candidate assembly, and flat-pack.

% BEGIN GRAPH POOL MEMORY
\subsection{Graph-Pool Memory}
Fig.~\ref{fig:graphpool}(d) reports graph-pool reserved memory for the evaluated
TP1 configurations with at most 32 active requests and memory fraction 0.85.
After initialization, we sum allocator segment sizes across distinct
nondefault graph pools, including unused reserved space.
DScale retains 832 MiB on Qwen3-8B and 724 MiB on Qwen3-4B.
This is 20.2\% and 11.3\% below DFlash, 27.1\% and 26.1\% below DSpark,
and 42.4\% and 36.5\% below Domino, respectively.

% END GRAPH POOL MEMORY
\section{Related Work}\label{sec:related}

\textbf{Speculative decoding.} Medusa and EAGLE use draft trees and target
features~\cite{cai2024medusa,li2024eagle,li2024eagle2,li2025eagle3}.
Lookahead uses Jacobi n-grams~\cite{fu2024lookahead}, LayerSkip uses early
exit~\cite{elhoushi2024layerskip}, and SpecInfer batches draft-tree
verification~\cite{miao2024specinfer}. DSpark and Domino combine parallel
backbones with token-dependent sequential correction~\cite{dspark,domino}.
Block diffusion generates tokens in parallel within a block~\cite{arriola2025block},
while DFlash emits its candidate block in one draft forward~\cite{chen2026dflash}.
SpecDec++ and AdaEDL stop an autoregressive drafter using predicted
acceptance or an entropy-based bound~\cite{specdecpp,adaedl}.
Instead, DScale retains full-block drafting and allocates reduced
verification-graph capacity among already-generated candidate prefixes
across requests, without token-by-token draft stopping.
Table~\ref{tab:related-capabilities} compares these serving capabilities.

\textbf{KV cache optimization.} PagedAttention reduces KV-memory fragmentation
and supports cache sharing~\cite{kwon2023efficient}.
SGLang's RadixAttention reuses common-prefix caches across requests~\cite{sglang}.
H$_2$O retains heavy-hitter and recent tokens while evicting other KV
entries~\cite{zhang2024h2o}. These methods optimize stored context, while
DScale reduces the candidate verification work added at each decode step.

\textbf{Serving scheduling and resource adaptation.} Orca combines
iteration-level scheduling with selective batching, while Sarathi-Serve and
DeepSpeed-FastGen schedule mixed prefill and decode workloads
~\cite{orca2022,agrawal2024sarathi,holmes2024deepspeedfastgen}.
DistServe separates the two phases across GPUs, and DOPD dynamically reallocates
their instances~\cite{zhong2024distserve,liao2026dopd}.
TightLLM reduces offloading overhead through adaptive KV recomputation and
cross-batch weight loading, whereas BrownoutServe merges experts and applies
brownout for bursty MoE workloads, trading accuracy for SLO attainment
~\cite{hu2025tightllm,hu2026brownoutserve}.
DScale instead preserves target computation while allocating per-request
verification budgets within reusable GPU graphs.

\textbf{Operator and attention optimization.} FlashDecoding++Next improves
inference execution through asynchronous softmax, shape-aware flat GEMM
optimization, and activation-buffer reuse~\cite{dai2025flashdecodingnext}.
MInference assigns sparse patterns to attention heads and constructs
input-dependent sparse indices to accelerate long-context
prefill~\cite{jiang2024minference}.
FlexPrefill adapts sparse patterns and computation budgets to each input
and attention head~\cite{lai2025flexprefill}.
These sparse methods select attention connections.
DScale jointly adapts query-row tiles, candidate budgets, and captured-graph
interfaces for short verification, preserving prefill tiles and full-length drafting.

\section{Conclusion}\label{sec:conclusion}

DScale combines path-aware tiles, score-guided half-capacity verification,
and fixed-address graphs with dynamic request boundaries.
It preserves the drafter's architecture, weights, and full draft length,
training only a small predictor without confidence calibration or hardware
speed-curve preparation. On A100 with TP1, two target models, four datasets,
and concurrency 8--32, geometric-mean throughput improves by 43.9--48.8\%
over DFlash, 22.2--37.7\% over DSpark, and 24.4--32.0\% over Domino,
with lower request latency. Ablations, accepted-token retention analysis,
and GPU profiling quantify the three mechanisms' gains and costs.
We retain the selected DFlash drafters' 16-slot blocks and leave other block
widths for future adaptation and evaluation.

\section*{Acknowledgment}

This work is supported by the National Key R\&D Program of China
(No. 2026YFE0199800), National Natural Science Foundation of China under
Grant 62572462, Guangdong Science and Technology Cooperation Project
(No. 2025A0505020065), Guangdong Basic and Applied Basic Research Foundation
(No. 2024A1515010251), Key Research and Development and Technology Transfer
Program of Inner Mongolia Autonomous Region (2025YFHH0110) and Shenzhen
Science and Technology Program under Grants JCYJ20240813155810014 and
ZDYJ20251211121533004.

\bibliographystyle{IEEEtran}
\bibliography{references}

\begin{IEEEbiography}[{\includegraphics[width=1in,height=1.25in,clip,keepaspectratio]{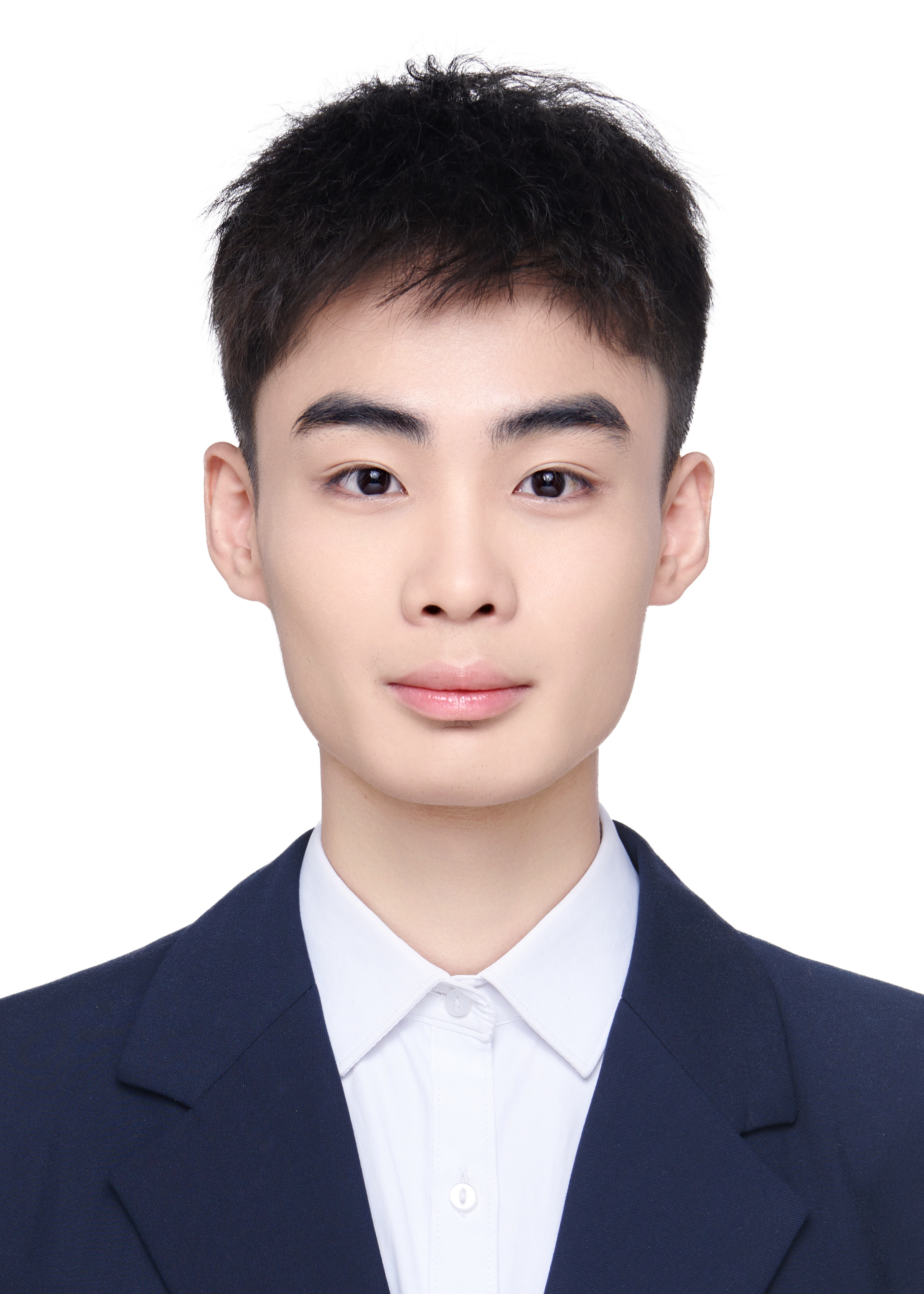}}]{Rongjian Chen}
received the bachelor's degree from Guangdong University of
Finance and Economics. He is currently working toward the master's degree with the
Shenzhen Institutes of Advanced Technology, Chinese Academy of Sciences. His
main research interests include large language models inference optimization
and system management.
\end{IEEEbiography}

\begin{IEEEbiography}[{\includegraphics[width=1in,height=1.25in,clip,keepaspectratio]{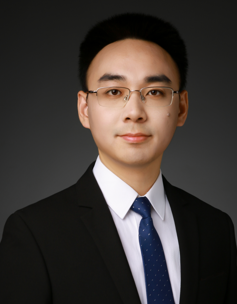}}]{Minxian Xu}
(Senior Member, IEEE) received the PhD degree from the University
of Melbourne, in 2019. He is currently an associate professor with the
Shenzhen Institutes of Advanced Technology, Chinese Academy of Sciences. His
research interests include resource scheduling and optimization in cloud
computing. He has co-authored 90+ peer-reviewed papers published in prominent
international journals and conferences. His PhD thesis was awarded the 2019
IEEE TCSC Outstanding PhD Dissertation Award. He was also awarded the 2023
IEEE TCSC Award for Excellence (Early Career Award).
\end{IEEEbiography}

\begin{IEEEbiography}[{\includegraphics[width=1in,height=1.25in,clip,keepaspectratio]{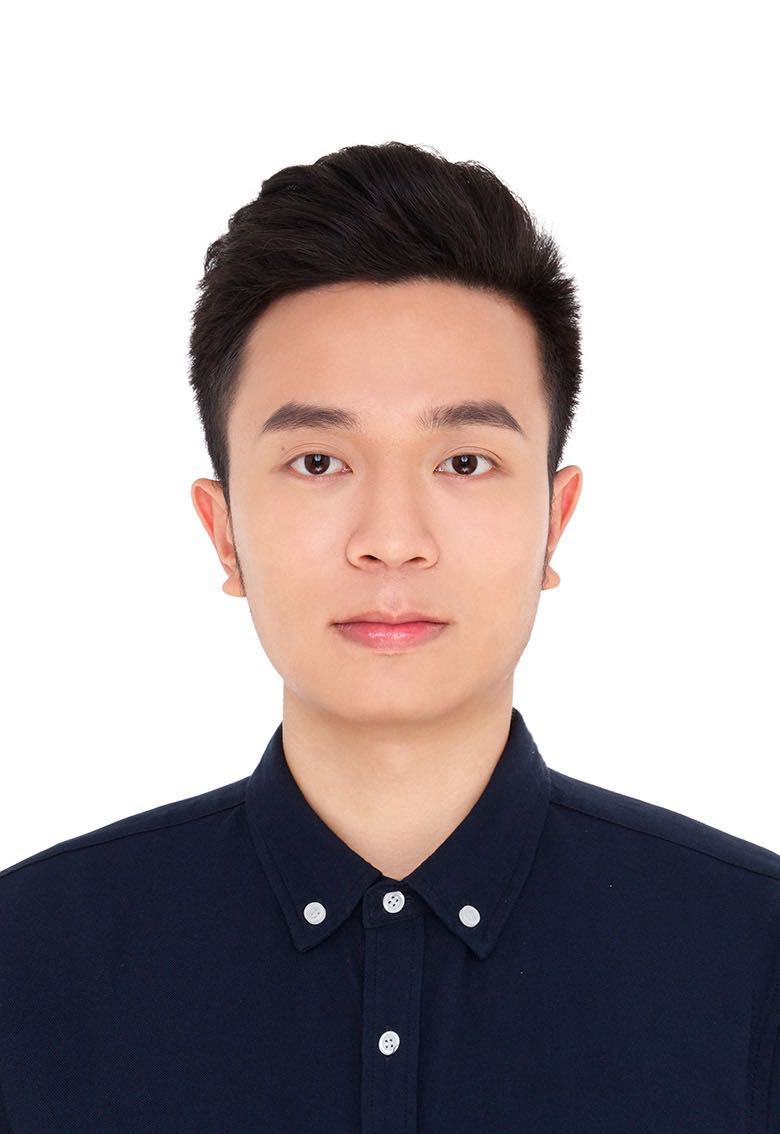}}]{Zhengxin Fang}
(Graduate Student Member, IEEE) received his bachelor and master degrees at
South China Normal University and Harbin Institute of Technology, in 2019 and
2021, respectively. He is currently pursuing the Ph.D. degree in Computer Science
at Victoria University of Wellington, New Zealand. His research interests include
cloud computing, microservice resource allocation, evolutionary computation,
graph neural networks, reinforcement learning and large language model assisted
optimization.
\end{IEEEbiography}

\begin{IEEEbiography}[{\includegraphics[width=1in,height=1.25in,clip,keepaspectratio]{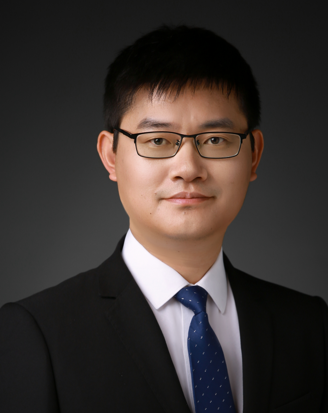}}]{Kejiang Ye}
(Senior Member, IEEE) received the BSc and PhD degrees from
Zhejiang University, in 2008 and 2013, respectively. He was also a joint PhD
student with the University of Sydney from 2012 to 2013. After graduation, he
works as post-doc researcher with Carnegie Mellon University from 2014 to
2015 and Wayne State University from 2015 to 2016. He is currently a
professor with the Shenzhen Institutes of Advanced Technology, Chinese Academy
of Sciences. His research interests focus on the performance, energy, and
reliability of cloud computing and network systems.
\end{IEEEbiography}

\begin{IEEEbiography}[{\includegraphics[width=1in,height=1.25in,clip,keepaspectratio]{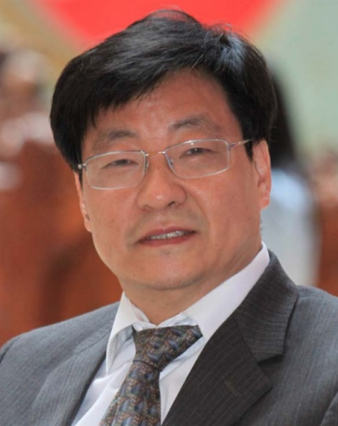}}]{Chengzhong Xu}
(Fellow, IEEE) received the PhD degree in computer science and
engineering from the University of Hong Kong, in 1993. He is with the
Institute of AI and Brain Sciences and the Department of Computer Science,
University of Macau. He published two research
monographs and more than 300 peer-reviewed papers in journals and conference
proceedings. His papers received about 17K citations with an H-index of 72.
His main research interests lie in parallel and distributed computing and cloud
computing.
\end{IEEEbiography}

\end{document}